\documentclass[runningheads]{llncs}

\usepackage{tikz}
\usepackage{amsmath}
\usepackage{ulem} 
\usepackage{xspace} 
\usepackage{float}
\usepackage{graphicx}

\newcommand{\highlight}[1]{\vspace{.05in}\noindent{\bf #1}:}

\newcommand{\etal}{\textit{et al}.}

\usepackage{wrapfig}
\usepackage{etex}
\usepackage{enumitem,kantlipsum}
\usepackage{epsfig,endnotes}
\usepackage{verbatim}
\usepackage[font=footnotesize,labelfont=bf]{caption}
\usepackage{subcaption}
\usepackage{amsfonts}
\usepackage{tabu}
\usepackage{color}
\usepackage{soul}
\usepackage{mathtools}
\usepackage{algorithm}
\usepackage{algpseudocode}
\usepackage{comment}
\usepackage{subscript}
\usepackage[version=3]{mhchem}
\usepackage{setspace}
\usepackage{multirow}
\usepackage{rotating}
\usepackage{syntax}
\usepackage{comment}
\usepackage{graphics}
\usepackage{colortbl}
\usepackage{tikzsymbols}

\usepackage{cite}[compress,sort]
\usepackage{url}
\usepackage{hyperref}

\hypersetup{draft}

\begin{document}

\date{}

\title{Security Issues and Challenges in Service Meshes -- An Extended Study}



\author{\rm Dalton A.\ Hahn
\and 
\rm Drew Davidson
\and 
\rm Alexandru G.\ Bardas
\institute{\textit{EECS Department, ITTC\\ University of Kansas} \\ Lawrence, KS USA\\ \{daltonhahn, drewdavidson, alexbardas\}@ku.edu}
}

\maketitle
\vspace{-0.3in}

\begin{abstract}
Service meshes have emerged as an attractive DevOps solution for collecting, managing, and coordinating microservice deployments. 
However, current service meshes leave fundamental security mechanisms missing or incomplete. 
The security burden means service meshes may actually cause additional workload
and overhead for administrators over traditional monolithic systems.
By assessing the effectiveness and practicality of service mesh tools, this
work provides necessary insights into the available security of service meshes.
We evaluate service meshes from two perspectives: skilled system administrators (who deploy optimal configurations of available security mechanisms) and default configurations.
Under these two models, we consider a comprehensive set of adversarial scenarios and uncover important design flaws 
with contradicting goals,  
as well as the limitations and challenges encountered in employing service mesh tools for operational environments. 
\end{abstract}

\section{Introduction}
\label{sec:intro}
The widespread enthusiasm of large enterprises for \textit{microservice} system architectures~\cite{balalaie16}, 
where many lightweight containers are managed and deployed via automation tools~\cite{docker20}, 
is unmatched by an evaluation of their security.
A number of academic works have examined the security of individual 
containers~\cite{combeDocker,martin18,enckDocker17}. However, \textit{service meshes} 
that manage microservice \textit{clusters}, remain largely unstudied.
This work focuses on microservices, due to the promise of a more adaptable, and flexible 
large-scale system deployment structure~\cite{balalaie16,jamshidi18,pahl16,johannes15}. 
The design philosophy underlying microservices is to refactor monolithic applications into 
collections of distinct components that collaborate at scale~\cite{microservices20}.

Service meshes ease the complexity of managing microservice architectures by allowing the 
administrator to express the structure and relationships between services using configuration 
files~\cite{consul-whitepaper19,istio-concept}. In contrast to traditional means of 
structuring system components via documentation (or knowledge shared by senior system 
architects), service meshes explicitly indicate dependencies between components. 
State-of-art service mesh tools such as Consul~\cite{consul20}, Istio~\cite{istio20}, and 
Linkerdv2~\cite{linkerd20} launch collections of microservices automatically, ensuring 
that microservices are deployed once dependencies are available. 
Furthermore, these tools automate \textit{service discovery}, the process of locating and binding 
services together. Service discovery is a non-trivial process under the 
DevOps~\cite{balalaie16,chen18,aws-devops20} ideology to 
support a variety of flexible deployments. As such, service discovery uses a 
decentralized model where dependencies are satisfied dynamically.

The purpose of service mesh tools is to provide a layer of abstraction over microservice launch and service discovery.
In doing so, service mesh tools carry the responsibility of orchestrating 
deployments securely. 
The decentralized nature of service meshes requires special care to ensure that malicious actors do not overwhelm 
cluster formation. In addition to attack vectors that are new to the service mesh domain, traditional 
system compromise remains a possibility. The goal of our work is to assess how well modern tools meet these challenges.

In studying service mesh security, we discover that misconfiguration issues and lack of 
security mechanisms enable numerous attacks. We view these as consequences of design flaws in service mesh security.
When facing these attacks, service meshes either offer no defense or require significant 
manual intervention on the part of the system administrator. The latter effectively undermines 
a core goal of service meshes: ease of automation. As such, our study suggests that 
service meshes should be redesigned to support security best-practices.

Current practices such as infinite-lifetimes and shared encryption keys~\cite{consul-keys20} 
indicate that the design of service meshes has overlooked important security concerns.
Nonetheless, deployment of immature service meshes is growing in production 
environments~\cite{consul-logic-monitor20,consul-spaceflight20,consul-neofonie20,consul-bitbrains20,consul-lithium20,consul-bol20,consul-distil20}. 
Aside from maturity concerns, configuring and maintaining these tools may come at a high cost to administrators, 
sometimes even greater than their original workflow demanded. 
Moreover, the context-dependent scope and implementations of service meshes are so diverse that 
establishing a meaningful comparison between different tools is difficult. 

Despite the building importance of defending service meshes, we are unaware of
any systematic assessment of their security.
To the best of our knowledge, this paper presents the first 
study to specifically focus on existing security mechanisms in service meshes. 
Our assessment indicates although service mesh tools embrace known 
consensus protocols such as RAFT~\cite{raft15} or extend membership protocols such 
as SWIM~\cite{swim02}, their security implementations and maintenance mechanisms are incomplete, or even non-existent. 
Additionally, we discovered that even though service mesh tools advertise their security 
contributions, they are either not enabled by default, or are left to third-party tools to implement.  

\vspace{.05in}
\noindent Our contributions can be summarized as follows: 
\begin{itemize}[leftmargin=*]
    \vspace{-.1in}
    \item We present the first study to examine the security design and analyze the available security mechanisms 
    within current service meshes  
    
    \item We propose a relevant threat model to the service mesh domain and assess the effectiveness of existing tools to mitigate these threats

    \item We assess the impact and the effort of utilizing available security features in current service mesh tools 
\end{itemize}

\noindent The remainder of this paper is structured as follows: Section~\ref{sec:back} introduces background while Section~\ref{sec:methodology} covers our threat model and the experimental design.
Section~\ref{sec:evaluation} presents both our experimental evaluations (idealized in Section~\ref{sec:idealized} and 
default Section~\ref{sec:default}). 
Related work appears in Section~\ref{sec:rel}, Section~\ref{sec:future} includes potential future work and 
we present our concluding thoughts in Section~\ref{sec:conc}.

\section{Background}
\label{sec:back}

Microservice architectures consist of a complex web of narrowly-scoped, 
interacting services in place of a monolithic architecture.
This structure better enables incremental changes, resilience to cascading failures, and quicker 
update/release cycles~\cite{chen18,cncf20,singleton16} at the cost of complexity; it is a significant 
challenge to maintain synergy between services.  Additionally, systems such as Kubernetes~\cite{kubernetes20}
provide a framework to deploy, scale, and manage microservices quickly, magnifying the need to coordinate services.
Service meshes seek to address this gap between fast deployments and collaborating webs of microservices. 
In this section, we describe some of the enabling tools and design concepts that underlie service meshes.

\highlight{Service Discovery and Management}
Service meshes enable a service to be registered to a cluster, discovered dynamically by other
dependent services, and to have configuration state maintained.  We note that some alternative 
tools provide service discovery and management capabilities, but are not classified as service meshes. 
In particular, Alibaba's Nacos~\cite{nacos20}, uses the Domain Name System (DNS) to collect, register
and maintain a list of available services~\cite{service-discovery20}. The concepts of service 
discovery and registry, have been present in distributed computing since the 
Hadoop-era~\cite{zookeeper20}, but have seen a return to fulfill the needs of
connecting services in microservice architectures.

\begin{figure}[t]
\begin{center}
\includegraphics[width=4.7in]{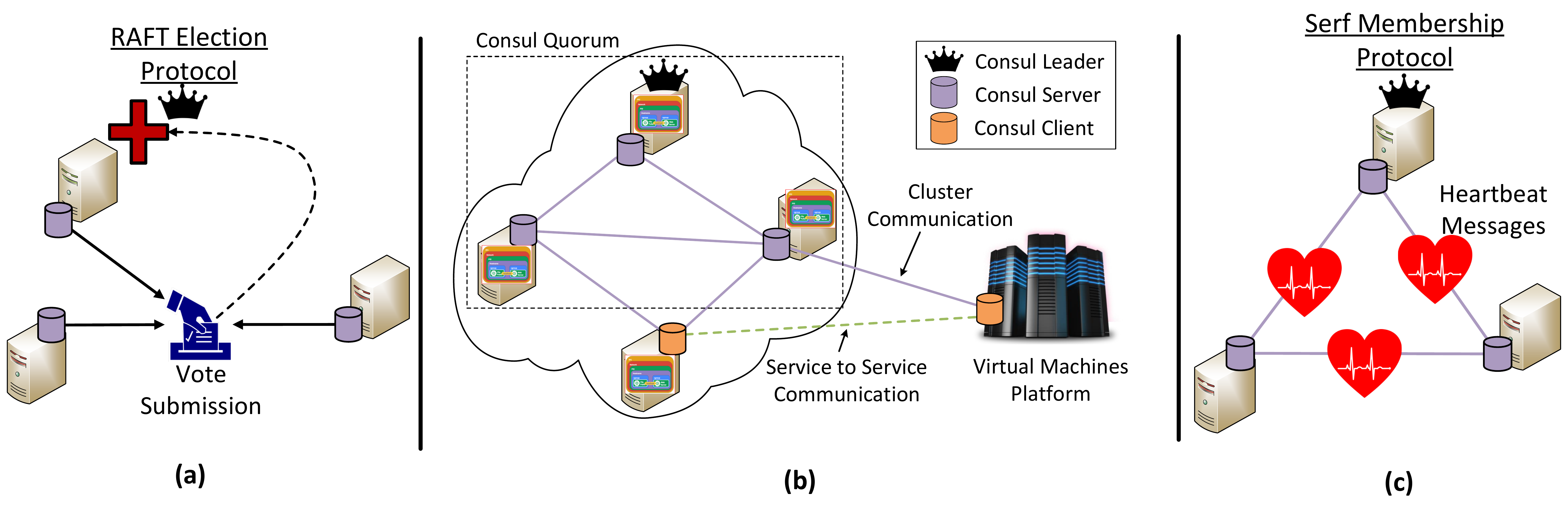}
\end{center}
\vspace{-.2in}
\caption{Model Consul Service Mesh -- Using Consul, the creation and operation of a model service mesh
are shown.  \textbf{(a).} RAFT elections occur periodically among Consul servers to determine cluster leadership.
\textbf{(b).} Proxies present on each node route cluster- and service-level
communications to nodes.  Proxies may be installed on a variety of platforms including virtualized,
containerized, and physical machines with little restriction on operating system~\cite{consul-compat20}.
\textbf{(c).} The Serf membership protocol occurs with high frequency to send
heartbeat messages among nodes to track health and membership.}
\label{fig:running-example}
\end{figure}

\highlight{Service Mesh Tools}
Consul, Istio, and Linkerdv2 are the current state-of-art service mesh tools with full, production-ready releases.  
A major cause of complexity in coordinating services is to determine cluster membership and node operation status. 
Consul implements the Serf~\cite{serf20} membership and node health protocol (an extension of SWIM~\cite{swim02}) and the RAFT 
consensus protocol. The basic process is illustrated in Figure~\ref{fig:running-example}.
Leveraging the cluster membership logic, Consul creates a membership hierarchy to organize the permissions that 
members of the cluster possess to take action within the cluster.
The Consul quorum is responsible for maintaining a consistent membership registry and holding cluster 
elections for the cluster permission hierarchy. 

\begin{figure}[t]
\begin{center}
\includegraphics[width=4.7in]{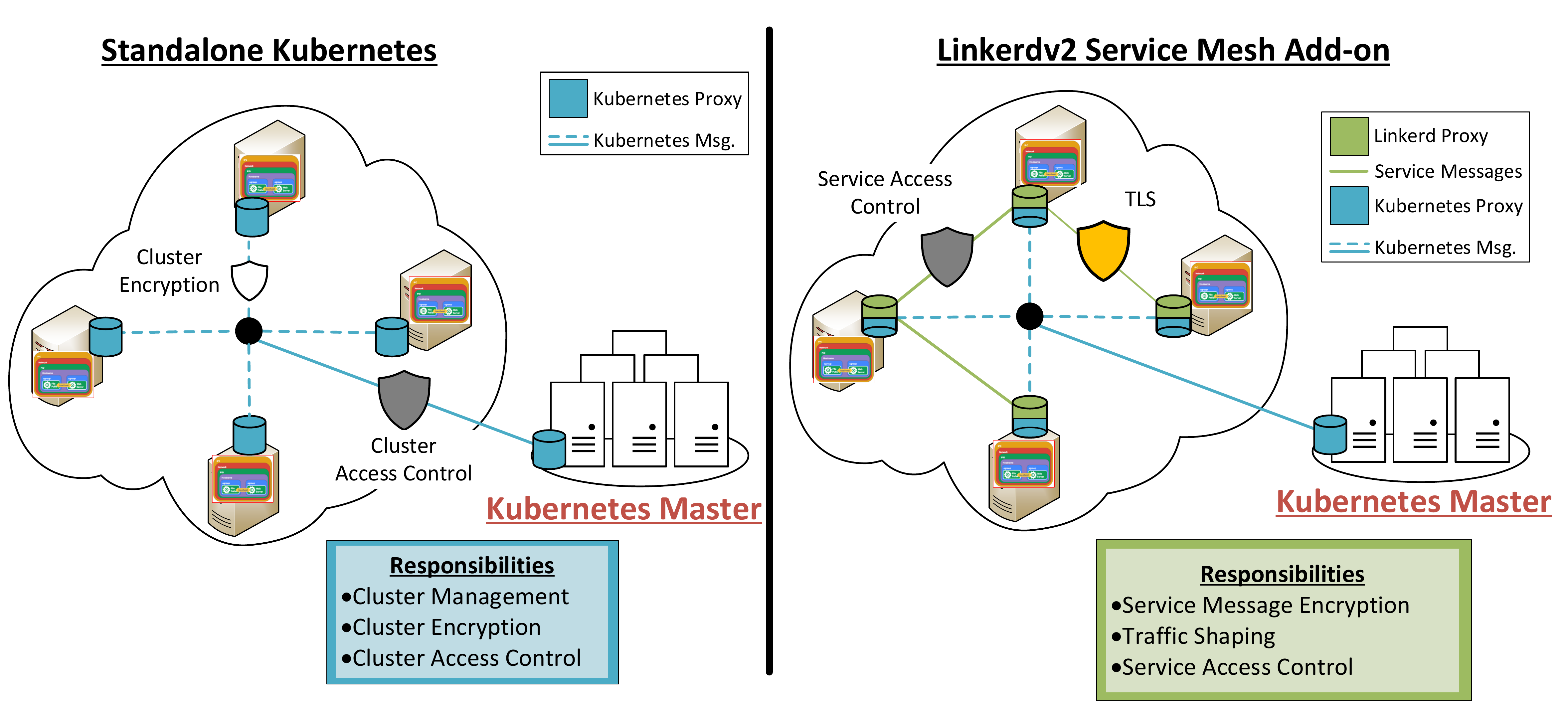}
\end{center}
\vspace{-.2in}
\caption{Linkerdv2 Service Mesh Structure -- A Linkerdv2 service mesh works {\it only} with Kubernetes. 
An established Kubernetes cluster provides the ``Cluster Access
Control'' security mechanism, but leaves ``Cluster Encryption'' to third-party tools.  Linkerdv2 provides
``Service Message Encryption'' through TLS and ``Service Access Control'' through the inherited Role-Based Access
Control from Kubernetes.}
\vspace{-0.1in}
\label{fig:linkerd-mesh}
\end{figure}

Istio and Linkerdv2 both require an underlying Kubernetes platform to provide cluster membership logic.
In contrast, installation of Consul is supported on a range of operating systems 
and architectures as well as virtualized and physical instances~\cite{consul-compat20}.
Without a previously created and configured Kubernetes cluster of \textit{pods}; collections of containers with 
shared resources~\cite{pods-kube20,lewis20}, Istio and Linkerdv2 are unable to provide
any of their promised features or security benefits. Figure~\ref{fig:linkerd-mesh} shows how an existing Kubernetes
infrastructure may be augmented by overlaying the Linkerdv2 service mesh on top.  However, by imposing the initial 
requirement of a properly installed, configured, and secured Kubernetes infrastructure, in addition to the overhead of 
configuring and maintaining the service mesh, Istio and Linkerdv2 demonstrate a higher burden 
on system administrators than that of Consul.
In contrast to Consul, Istio and Linkerdv2 do not maintain a hierarchical structure 
for permissions and state management, instead, they rely upon a star topology-like system where the Kubernetes master controls 
the cluster's pods either remotely, or locally, and sets the configuration and permissions of specific members within the cluster. 

\highlight{Service Mesh Security}
The paradigm shift from monolithic systems to microservice systems has caused a change from 
\textit{intra}-service issues to \textit{inter}-service issues.  This transitions the burden of 
security from within the operating system of a machine to across network connections.  
Issues previously addressable by trusted security measures within the operating system must 
now be addressed with network-level security measures. These issues include the need to protect 
cluster-level communications, service-level communications, and access permissions, both at the 
cluster-level as well as the service-level.  
Within monolithic applications, membership is addressed within software design and all software 
components are ``members'' of the larger system through the architectural design of the software. 
However, in microservices, the software application is divided and fragmented into functional 
components which must then be connected via a network medium, necessitating a network-level solution.  
Service-to-service communications within monolithic applications could be sent via RPC or to components 
listening on the machine's loopback address.  In contrast, the nature of microservice architectures 
and their involved entities (containers, virtual machines, or physical machines) require secure network 
communications to relay service-level messages. 

\section{Threat Model and Experimental Design}
\label{sec:methodology}
To evaluate the security of modern service mesh tools, we used Consul as a model for service
mesh design and implementation. We constructed a proof-of-concept environment using Consul 
to conduct our experiments. We consider the available security mechanisms 
for administrators and examine a deployment utilizing all available mechanisms as well 
as one using default configurations. Under these setups, we conduct a series of active 
attacks and report our results. We also present a comparison of available and default security mechanisms within
Istio and Linkerdv2 and provide our findings. We utilize these findings to frame a discussion of the shortcomings 
and overhead system administrators should expect when attempting to secure service mesh 
clusters within their infrastructure. 

Consul provides a meaningful representation of service meshes and the maturity of these tools.
Of the current state-of-art service meshes, Consul is the most feature-rich and flexible tool available
in this domain.  As mentioned previously, Consul can be used with any other tools or forms of virtualization such as
containers or virtual machines whereas Istio and Linkerdv2 are dependent upon an underlying Kubernetes implementation
to provide necessary features for the mesh.  Additionally, as of the writing of this work, Consul appears to be the most 
actively developed tool, enjoying the largest number of GitHub contributors (594) of the tools we encountered, and a comparable
number of GitHub repository stars to the runner-up tool, Istio~\cite{consul-github-metrics20,istio-github-metrics20}).

\highlight{Threat Model}
\label{sec:threat}
The threat model we employ in this work considers common
attacker goals of disruption of services and exfiltration
of sensitive data.  However, we also consider adversarial
targets that are unique to the service mesh domain.
For example, an attacker may often desire to infiltrate the cluster and gain privilege rather than 
destroying the functionality of a system. By infiltrating the cluster, the attacker may inject malicious 
service configurations to possibly redirect benign service requests to externally controlled endpoints. 
Additionally, attackers may wish to maintain a position of strength within the cluster's leadership to gain capabilities 
to register malicious services within the cluster, remove defender-controlled resources, 
and acquire access to more computing resources for larger scale attacks, such as those executed by botnets. 
In Table~\ref{tab:attack-table}, we denote these high-level goals as \textbf{D}isruption, \textbf{M}anipulation, 
and \textbf{T}akeover for disruption to services and cluster activities, tampering of sensitive data via 
manipulation, and gaining privilege through service mesh takeover, respectively.

\begin{table*}[t]
\resizebox{\textwidth}{!}{
 \begin{tabular}{l | c | c | c | c | c } 

   & Datacenter Label  & UDP  & ACLs & TLS  & All Mechanisms Combined \\ 
   & as a Secret & Encryption & (Access Control Lists) & Encryption & (Datacenter Label, UDP, ACLs, TLS) \\
 \hline\hline
 Unprivileged Threat & \cellcolor[HTML]{f73c00}D M T & --- & \cellcolor[HTML]{ffcfcf}D 
 & --- & --- \\ \hline
 
 Client Compromise & \cellcolor[HTML]{f73c00}D M T & \cellcolor[HTML]{f73c00}D M T & 
 \cellcolor[HTML]{ffcfcf}D & \cellcolor[HTML]{ffcfcf}M & --- \\\hline
 
 Server Compromise & \cellcolor[HTML]{f73c00}D M T & \cellcolor[HTML]{f73c00}D M T & 
 \cellcolor[HTML]{ffcfcf}D & \cellcolor[HTML]{ffcfcf}M  & --- \\\hline
 
 Leader Compromise & \cellcolor[HTML]{f73c00}D M T & \cellcolor[HTML]{f73c00}D M T & 
 \cellcolor[HTML]{f73c00}D M T & \cellcolor[HTML]{f73c00}D M T & \cellcolor[HTML]{f73c00}D M T \\  \hline\hline 
 \end{tabular}
 }
 \vspace{.05in}
 \caption{Adversarial Goals on a Consul Deployment -- Presents experimental results of 
 achieved adversarial goals on a properly configured  Consul service mesh deployment.
 \textbf{D}isruption: Interruption to service availability.  
 \textbf{M}anipulation: Infiltration or exfiltration of data to cluster.
 \textbf{T}akeover: Adversary assumes the leadership position in cluster.}
 \vspace{-0.3in}
 \label{tab:attack-table}
\end{table*}

\highlight{Experimental Setup} 
We deployed our model cluster upon a Dell R540 
server configured with 128 GB of RAM, Xeon Gold 5117 processor, and 10 TB of SSD storage. 
We believe this hardware to be comparable to what would be utilized in production environments, 
both in on-site and remote, cloud datacenters. 

The proof-of-concept Consul service mesh consists of an initial leader node or ``bootstrapper'' 
responsible for initializing the cluster and connecting the initial nodes. 
Alongside the leader node are two server nodes, forming the quorum, and a singular client node.
Using Figure~\ref{fig:running-example} as our model, we manually deployed and configured these four Consul nodes
(one leader, two servers, and one client node). We utilize only one client node due to the equivalent
functionality of subsequent clients.
Due to the architecture of Consul service mesh clusters, it is recommended 
to have 3 nodes acting as servers (one leader node and 2 server nodes) to manage 
the quorum and maintain the cluster state and log files~\cite{consul-consensus20}. 
Nodes are the main structural components of service mesh clusters, hosting ephemeral or long-lived microservices
on permanent, virtual, or physical instances.

Once installed and configured, we run the Consul application to connect and register components
of the cluster and allow the leadership quorum to be established. Next, services were connected and the 
service mesh began routing cluster requests to their appropriate destinations.
Once services were configured, in order to enable all available security mechanisms for the cluster, the datacenter 
label was distributed, UDP encryption key created and distributed,
proper node-level and service-level access permissions set, and TLS certificates created and distributed appropriately.

\section{Evaluation of Modern Service Meshes}
\label{sec:evaluation}
In order to examine all aspects of the security of service meshes, we perform two,
parallel evaluations of the studied tools.  Under one, we consider an administrator
with deep knowledge of the employed tool and its available security mechanisms.  We
classify this as an ``idealized'' defense scenario.  Second, we study the employed
tools under their default configurations and report our findings to demonstrate
the significant burden placed upon administrators to properly enable and configure
the security mechanisms, or lack thereof, of these tools.

\subsection{Evaluation of the Idealized Service Mesh Defender}
\label{sec:idealized}
\vspace{-.1in}

Under our initial adversarial evaluation, we consider an administrator with deep 
knowledge of the available security mechanisms and their correct configuration. 
As such, an administrator can leverage these security mechanisms to their greatest potential.
To study how varying degrees of attacker strength can affect the level of compromise under 
these security mechanisms, we position the adversary at different levels of initial compromise.
The lowest initial power we consider an attacker to have is that of an ``Unprivileged Adversary''
who has not yet compromised any node within the cluster.  The highest initial level of power we consider
is that of ``Leader Compromise'' where an adversary has the preliminary position of a node considered
to be the leader of the Consul quorum.
Under the assumption of a knowledgeable administrator and the preconditions placed upon
the adversary, we evaluate the experimental results and provide our assessment. 

\begin{figure}[t]
\includegraphics[width=\linewidth]{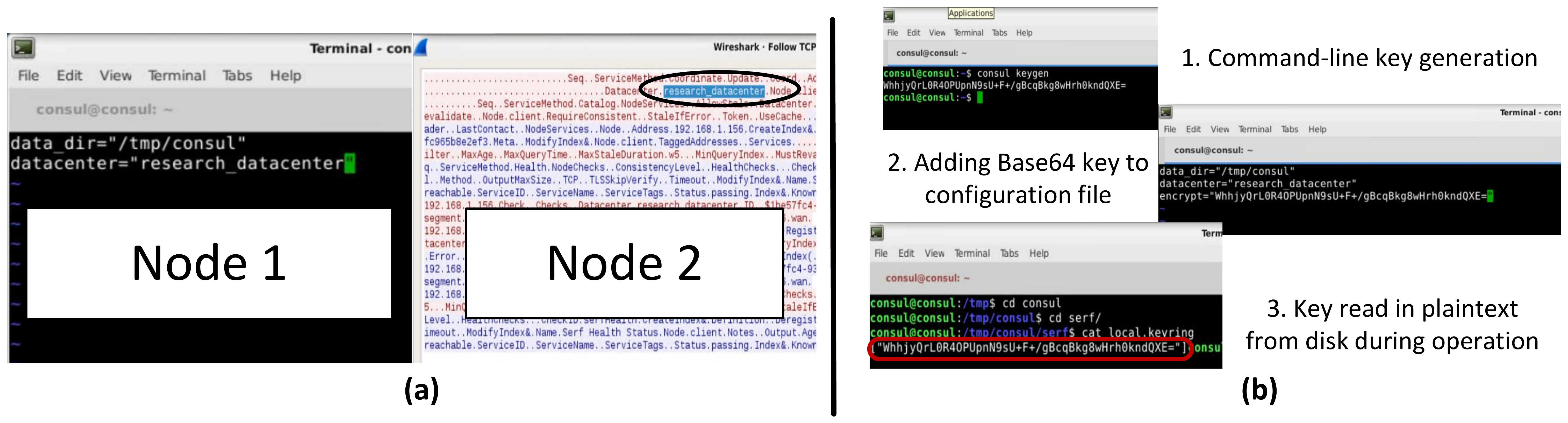}
\vspace{-.25in}
\caption{Plaintext UDP Key Storage and Plaintext Transmission of Datacenter Label --  
\textbf{(a).} Due to the plaintext
transmission of the datacenter label over the network during the node join process, a packet capture
software such as Wireshark~\cite{wireshark20} can be used to extract the datacenter label and illegitimately join
a target cluster.
\textbf{(b).} Shows the creation of a Base64 UDP encryption key using the built-in Consul
key generator and the subsequent plaintext storage of the generated key. }
\label{fig:datacenter-label}
\end{figure}

\highlight{Consul -- Datacenter Label as a Secret}
The first means of potential defense we consider within our proof-of-concept Consul service mesh 
is the datacenter label.  We consider this a potential security mechanism due to the fact that if
a prospective cluster node is configured with a datacenter label that differs from the target cluster, 
the prospective node will be denied membership to the cluster. In order to enable this configuration, 
a system administrator would be required to generate a datacenter label and distribute it among 
all members of the cluster. Once configured among the nodes, cluster 
initialization may commence. 

As shown in Table~\ref{tab:attack-table}, under all adversarial scenarios, using strictly ``Datacenter
Label as a Secret'' is insufficient in thwarting attacks against the cluster.  Specifically, when using 
datacenter label alone, communication messages are exchanged in plaintext between nodes of the cluster.  Figure~\ref{fig:datacenter-label}(a) shows how a captured communication packet between two Consul nodes exposes
this configuration detail to potential adversaries.  Due to the realistic
possibility of an adversary to capture a single packet exchanged between the nodes of the cluster, they may 
extract the datacenter label from the packet.
The malicious join operation is, subsequently, made possible 
due to a lack of UDP or TLS encryption among the nodes of the cluster. Once a member of the cluster, lack of any 
additional security measures concedes the ability for an adversary to conduct any 
action and command within the cluster without restraint.  Due to this, all high-level attacker goals can be achieved.

\highlight{Consul -- UDP Message Encryption}
Next, we consider the Consul service mesh deployed using UDP message encryption as the sole mechanism of defense.
As shown in Table~\ref{tab:attack-table}, enabling UDP message encryption thwarts an unprivileged adversary from
achieving any of their goals, but fails to provide protection under compromise of any cluster members. 
By enabling UDP encryption, the adversarial joins previously possible are prevented because an attacker  
is unable to decrypt packets from the legitimate nodes. 

All nodes within a Consul service mesh share the \underline{same} encryption key. To exacerbate this concern, Consul, 
as of the writing of this work, fails to provide any means of key revocation or rotation. Figure~\ref{fig:datacenter-label}(b)
shows the plaintext storage of the UDP encryption on disk of all Consul nodes, further demonstrating how the implementation 
of this security mechanism has been patched into the software, rather than accounted for in system design.
In order to provide key rotation within the cluster, even through a separate ``recovery'' mechanism such as an SSH~\cite{ssh} 
session, all nodes must be stopped, configurations adjusted, and the cluster recreated. Lack of key rotation support 
deepens the potential damage of key exposure and burden placed upon administrators. The long-term implications
to maintain a cluster is clearly taxing upon its users.  While the managed services of the cluster may be 
transient and possibly
short-lived, the underlying service mesh infrastructure is intended to be long-living.  Therefore, support for
key rotation capabilities is vital for managing and maintaining a secure service mesh architecture.

During the lifetime of a Consul cluster secured with solely UDP message encryption, should a single node
be compromised, the entirety of the service mesh is compromised.  Due to the shared key among all nodes, 
a single compromise allows an adversary to replicate the key among malicious nodes and join the cluster or exfiltrate the shared key to other adversarial nodes.
Once becoming a member of the cluster, an adversary is, once again, able to perform any action within the 
cluster without restriction. Such actions include creating new service configurations,
removing the current leader node, and reading secrets from the key/value storage system. 

\highlight{Consul -- ACLs}
As shown in Table~\ref{tab:attack-table}, Access Control Lists (ACLs) are highly effective at thwarting 
the adversarial goals of manipulation and takeover within the cluster. However, ACLs prove futile against 
disruption of cluster activities and service availability. In order for a system administrator to enable 
ACLs as a defense mechanism, extensive permission policies must be created and access tokens exchanged using 
a third-party, secure channel such as SSH.  With a lack of support for distributing security objects safely within
Consul itself, the implementation of ACLs, and subsequently the policies and tokens generated, demonstrates that security
mechanisms within service meshes have been ``bolted-on'' to existing software, rather than incorporated into system design.

\begin{figure}[t]
\centering
\includegraphics[width=4.7in]{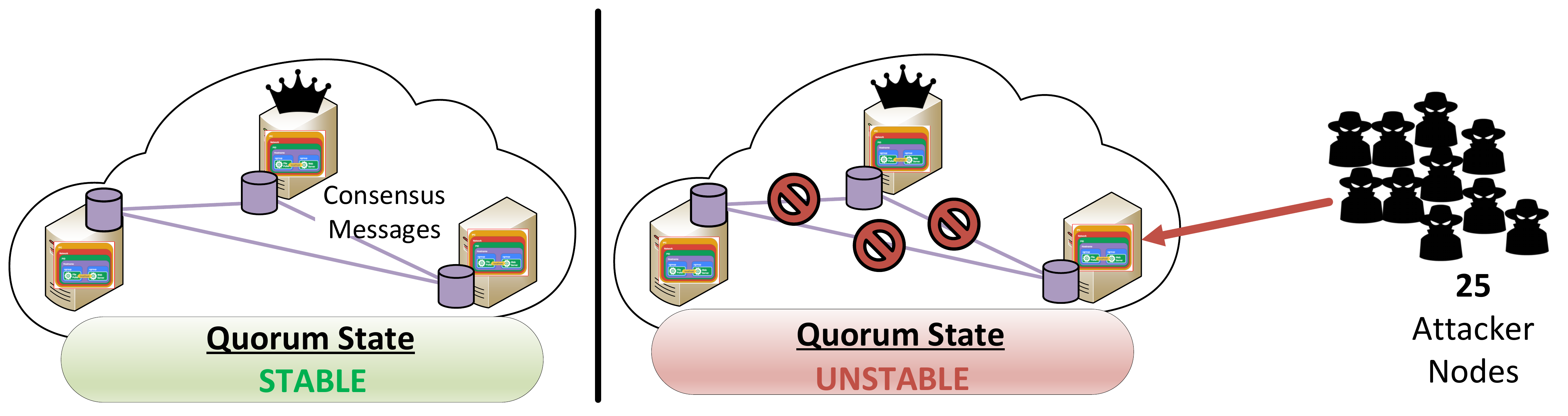}
\vspace{-0.05in}
\caption{Disruption Attack -- Depiction of a 3 node Consul server quorum.
An adversary can flood the cluster with malicious
nodes with the ``server'' flag to overwhelm the RAFT consensus
protocol within the Consul service mesh.  We find that 
less than 25 attacker nodes were required to disrupt cluster operations
and create instability within the Consul quorum.}
\vspace{-0.15in}
\label{fig:quorum-overwhelm}
\end{figure}

In order to secure the simple, four node service mesh used for our evaluation, as advised by the Consul 
tutorials~\cite{consul-acls20}, an administrator would need to generate unique access policies, generate tokens, 
and distribute and assign the generated tokens to proper recipients.  All of these actions must be conducted from
the single leader node due to the advised ``operator-only'' policy. Under the ``operator-only'' policy, 
permissions to edit the ACLs are restricted to the leader, meaning a singular node is responsible for all creation 
and distribution of policy materials. In direct contrast to the decentralized, distributed nature of the service mesh,
the security structure implemented has been consolidated to a single point of control, the Consul leader.  
Augmenting the burden placed upon system administrators, the current 
implementation of Consul ACLs have no token rotation policy in place. Therefore, either the created access tokens 
within the cluster will exist for the lifetime of the cluster, or are revoked after a period of time, but with no means of 
redistributing fresh tokens to nodes. 

Aside from the difficulty encountered by administrators to simply establish a secure permissions structure, there 
is also the potential for adversarial action. When ACLs are the sole mechanism of defense for a Consul service 
mesh cluster, they prove ineffective at mitigating adversarial disruption efforts. Due to implementation of 
processing access control policies within the service mesh, unauthorized messages must be confirmed as illegitimate 
by the cluster. Using around 25 adversarial nodes, we were able to disrupt operations within the service 
mesh by overwhelming the consensus protocol, see Figure~\ref{fig:quorum-overwhelm}.  
By configuring the malicious nodes with the Consul server flag, the malicious nodes were able to 
generate an overwhelming number of false consensus messages within the cluster, preventing the benign nodes 
from operating effectively and maintaining control of the cluster. The joining of malicious nodes is 
made possible due to a lack of UDP or TLS encryption within the cluster, exposing a lack of synchronization among
the security mechanisms available.  
While the malicious nodes may be prevented from taking action due to lack of proper permissions, they may 
leverage scale to overwhelm the cluster. Due to the need to verify the legitimacy of the packets, computation 
is expended by the legitimate members of the cluster and may subsequently overwhelm them. 
By doing this, an adversary may halt the availability of the quorum and prevent an administrator from 
creating new policies or access tokens for the cluster. 
Despite having no legitimate access to the cluster, using this process, an adversary may still achieve 
their goal of disruption within a cluster when ACLs are utilized as the sole means of defense.

\begin{figure}[t]
\center
\includegraphics[width=4in]{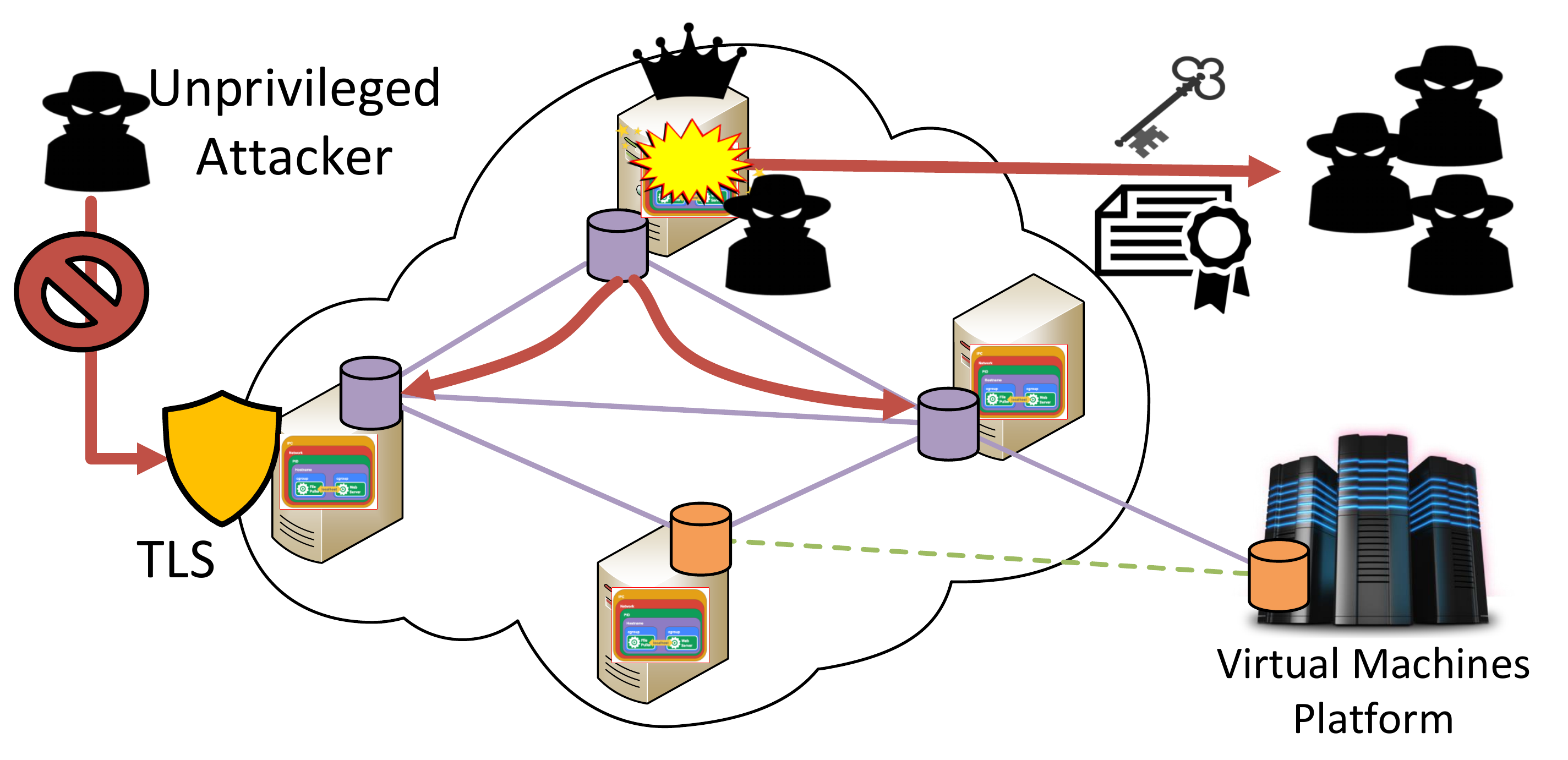}
\caption{TLS Message Encryption -- Encrypting service traffic with TLS prevents an
unprivileged attacker influencing the cluster. However, a leader node compromise 
allows an adversary to generate malicious TLS key pairs and exfiltrate them to other 
adversary nodes. Once additional adversaries join they may
join the quorum and cast votes due to their server-permissioned certificates.}
\vspace{-0.2in}
\label{fig:tls-encryption}
\end{figure}

\highlight{Consul -- TLS Message Encryption}
In order to protect service-level communication within the Consul cluster, a system administrator may enable 
TLS message encryption.  To provide nodes the capability to sign messages, they must first have signed certificates
from the certificate authority.  When constructing the service mesh, the administrator would create a certificate
authority from one of the server nodes of the cluster.  Afterwards, the certificate authority is responsible for
generating server certificates and client certificates as well.  Distribution of certificates must be completed 
before the cluster may be constructed and connected. Additionally, in order to ensure that client nodes may not alter 
their configuration and obtain server permissions within the cluster, the ``verify\_server\_hostname'' flag must 
be set in all node configurations.

There are many scenarios under which a privileged adversary may still accomplish some of their goals once 
the cluster is constructed and certificates distributed. However, by enabling TLS encryption, the unprivileged 
adversary is unable to maliciously join the cluster, preventing any goals from being achieved in this case. 
Despite this, there are no protections for the key/value storage system should a client node be compromised 
due to TLS being the sole mechanism of defense. Accessing the key/value
storage allows an adversary to manipulate configurations or secrets stored within the cluster.  
Figure~\ref{fig:tls-encryption} shows how, should the leader node ever be compromised in the lifetime
of the cluster, an adversary may leverage the signing privileges of the certificate authority to generate 
illegitimate certificates and keys for malicious nodes. 

The implementation of the certificate hierarchy within Consul once again shows a disconnect between the desired
decentralized and distributed nature of service meshes with a centralized, consolidated security structure.  Within
Consul, the only node able to sign certificates of any privilege is the certificate authority (commonly created on the 
quorum leader node).  Additionally, this indicates that should the leader node ever be destroyed, unless the certificate 
authority key was replicated to other nodes, the cluster has lost the ability to sign new certificates, once again conflicting
with the flexibility goal of the DevOps ideology.  Lastly, a lack of revocation and rotation mechanisms 
within Consul itself necessitates a third-party tool such as HashiCorp's Vault~\cite{vault20} or SSH be used to 
distribute fresh certificates to nodes, which triggers the need for widespread edits to configuration files and application
restarts to transition the service mesh to a secure state once again.

\highlight{Consul -- All Mechanisms Combined}
By enabling and combining all available security mechanisms, Table~\ref{tab:attack-table} shows a clear 
improvement in mitigating adversarial goals. However, employing all mechanisms presents 
administrators with a daunting amount of manual configuration. Considering the cost required to establish 
a secure model example with trivial functionality, the requirements to successfully deploy and secure 
enterprise-level systems is unreasonable. Also, due to the implementation of the available security mechanisms, 
should the leader of the cluster ever be compromised across the lifetime of a cluster, all effort to construct 
the service mesh must be repeated in order to redeploy a secure service mesh state. 
By lacking necessary revocation and rotation mechanisms, Consul has limited the ability to
construct dynamic service mesh clusters that are resilient to compromise events. 
Due to this, the assumption that nodes will never be compromised has created dangerous situations for system 
security where a single compromised node may result in collapse of the entire cluster's security. 
Service mesh tools, while aiming to fill the niche of microservice architecture discovery, connection, 
and management, may, in fact, lead to substantial overhead for administrators 
who wish to deploy these tools in a secure fashion.

\highlight{Idealized Istio and Linkerdv2, with Kubernetes}
In contrast to the experimental setup of our proof-of-concept Consul service mesh, Istio and Linkerdv2 
both require an underlying Kubernetes infrastructure in order to be utilized.  
Under a skilled administrator model, we consider the creation, configuration, and security of Kubernetes 
to provide the necessary foundation for both Istio and Linkerdv2.  
Due to the dependent relationship between Istio and Linkerdv2 with Kubernetes, 
it is important to consider the mechanisms by which Istio and Linkerdv2 implement and 
deploy their security mechanisms and features.

In a similar fashion to Consul, Istio and Linkerdv2, when combined with Kubernetes, implement service-to-service communication 
encryption through TLS.  Within a Kubernetes cluster, sensitive data such as encryption keys and certificates may be transferred 
through a secure channel in the Kubernetes control plane referred to as ``Secret Volumes''~\cite{kubernetes-secrets20}. 
Istio and Linkerdv2 utilize secret volumes to transfer TLS certificates 
to the pods of a Kubernetes cluster~\cite{istio-tls-secrets20,linkerd-tls-secrets20}.  
Once the certificate has been successfully transferred, mutual TLS exchanges may be enabled within
Istio and Linkerdv2's service mesh configurations.
In contrast to Consul's implementation of ACLs, Istio and Linkerdv2 leverage the Kubernetes service registry to
provide the means for a similar access control system through Role-Based Access Control (RBAC) policies.  However, similarly to 
Consul, establishing meaningful and fine-grained RBAC requires extensive configuration on the part of a system administrator.
Istio and Linkerdv2 leverage RBAC, to secure service authorization and service-level permissions throughout the cluster. 
Utilizing the Kubernetes security capabilities, along with the security features that Istio and Linkerdv2 make available to 
administrators, namely service-to-service communication encryption and role-based access control to services, the combination 
of Kubernetes with either Istio or Linkerdv2 creates a comparative environment to a service mesh deployed using 
solely Consul.  This comparative service mesh environment provides a similar level of security as a properly 
configured and completely implemented set of security mechanisms available in Consul, however a user is locked 
into and required to use the Kubernetes platform, eliminating any choice of alternative.

\begin{figure}[t]
\centering
\includegraphics[width=4.7in]{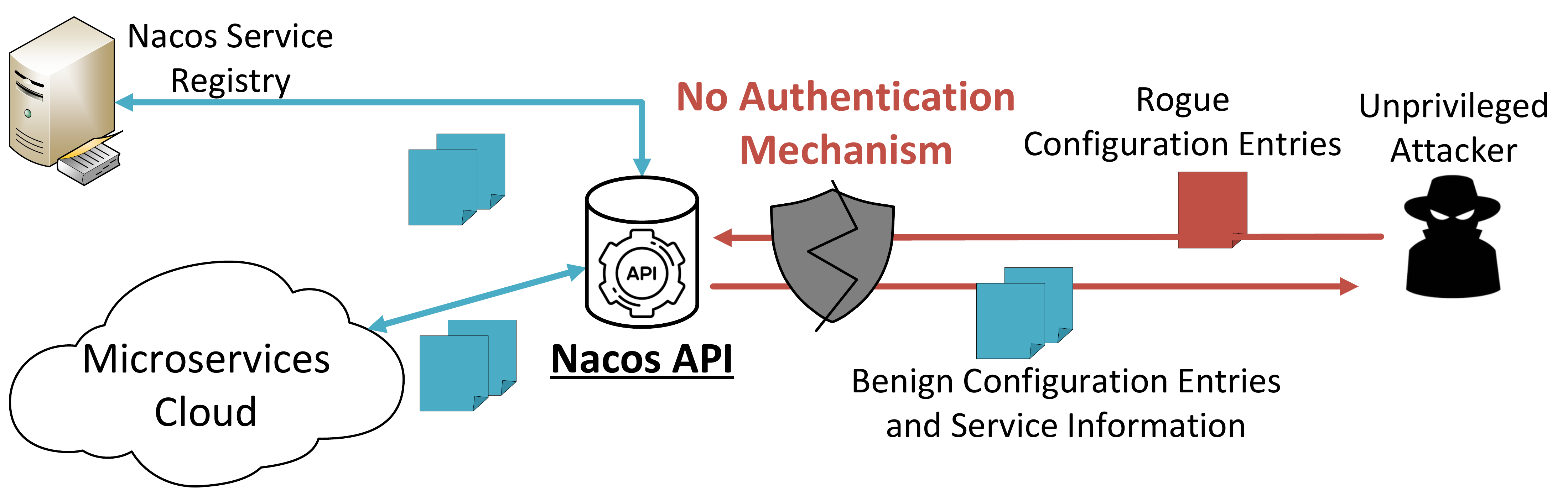}
\vspace{-0.05in}
\caption{Nacos Attack Scenario -- If an unprivileged attacker is able to 
locate a Nacos API server running on a publicly routable IP address, they may submit
any API requests to the service registry.  A lack of authentication mechanisms within
Nacos allows rogue configuration entries to be submitted to the registry or
benign service information and configurations to be read by the adversary.}
\label{fig:nacos}
\vspace{-0.2in}
\end{figure}

\highlight{Nacos}
While not directly considered a service mesh, Nacos provides many of the same features as the service meshes 
considered and has the ability to be configured in a way to accomplish many of the same high-level goals as service 
mesh tools. However, it is important to note that Nacos is technically a service discovery and management tool.  
As of the writing of this work, Nacos is in version 1.1.4, and is available for public use. However, Nacos
has very little, if any security mechanisms available to its users. As a service discovery and management tool, Nacos
attempts to coordinate, collect, and store information regarding the services available in a deployed microservice 
architecture. However, when using Nacos, there are no protections to the API made available to users. Due to a lack
of authentication, any Nacos server visible on public IP addresses is vulnerable to adversarial threats.  
Figure~\ref{fig:nacos} shows how such an attack could be executed against a public-facing Nacos instance. 
An attacker can take any number of actions against a Nacos cluster including registering its own IP address 
as an instance of the cluster, adding, modifying, or removing service entries from the service registry, 
and adding, modifying, or removing configuration entries from the registry. An additional consequence of the ability 
to view configurations of the cluster is the fact that credentials may be stored within configuration files in 
order to coordinate with service consumers the correct method of authentication. By releasing a publicly available version 
of the tool before 
any security considerations have been implemented, Nacos exposes its consumers to a range of attacks that they may 
not be aware of, while offering no means of protecting themselves using the functionality of the tool. 
In its current state, Nacos depends primarily upon external security mechanisms such as firewalls, subnetting, and other 
perimeter defenses.  However, in microservice architectures with many components distributed across cloud providers,
the boundaries of an enterprise network are difficult to discern.  Often, these boundaries are volatile and change over time.
Due to this, public access to a Nacos cluster can be found through scanning engines such as Shodan~\cite{shodan} 
and Censys~\cite{censys}.  Shodan and Censys both provide keyword searches within service banners as well as port-oriented 
searches.  Leveraging these capabilities, an adversary could potentially locate vulnerable Nacos instances by searching for 
its default port, or keywords that appear in the Nacos service banner.  

Additionally, varying regulations around the world
sometimes prevent or discourage the use of encrypted traffic within networks~\cite{lo01}.  Also, regulation pressures 
may impact design decisions, such as the choice to delay development of encryption mechanisms.
According to the Nacos Blog~\cite{nacos-roadmap}, permission control for configurations, permission control for services, 
encrypted storage of configuration files, and other security mechanisms are part of planned future software updates. 
Due to the skilled system administrator having no security mechanisms available to them, we consider the assessment of the 
default configuration to be the same as the skilled case.  Therefore, we will not include a default assessment 
of Nacos in Section~\ref{sec:default}.

\subsection{Evaluation of Default Service Mesh Configuration}
\label{sec:default}
\vspace{-.1in}

\begin{table}[t]
\scriptsize
\def\arraystretch{1.3}
\centering
\begin{tabular}{l | c | c | c || c | c | c}
{\multirow{2}{*}{Tool}} & {\multirow{2}{*}{Security Mechanism}}  & \begin{tabular}[c]{@{}c@{}}Available\end{tabular} & 
\begin{tabular}[c]{@{}c@{}}Enabled \end{tabular} & \begin{tabular}[c]{@{}c@{}}Default\end{tabular} & \begin{tabular}[c]{@{}c@{}}{\multirow{2}{*}{Revocation}}\end{tabular} & \begin{tabular}[c]{@{}c@{}}{\multirow{2}{*}{Redistribution}}\end{tabular}\\
& & in Tool? & by Default? & Lifetime & &\\
\hline\hline
\parbox[t]{2mm}{\multirow{4}{*}{\rotatebox[origin=c]{90}{Consul}}} & \begin{tabular}[c]{@{}l@{}}Cluster Message Encryption\end{tabular} & Yes & \cellcolor[HTML]{ffcfcf}No & \cellcolor[HTML]{ffcfcf}$\infty$ & \cellcolor[HTML]{ffcfcf}No & \cellcolor[HTML]{ffcfcf}No\\ \cline{2-7}
 & \begin{tabular}[c]{@{}l@{}}Service Message Encryption\end{tabular} & Yes & \cellcolor[HTML]{ffcfcf}No & 1 year & Yes & \cellcolor[HTML]{ffcfcf}No\\ \cline{2-7}
 & \begin{tabular}[c]{@{}l@{}}Cluster Access Control\end{tabular}     & Yes & \cellcolor[HTML]{ffcfcf}No & \cellcolor[HTML]{ffcfcf}$\infty$ & Yes & \cellcolor[HTML]{ffcfcf}No\\ \cline{2-7}
 & \begin{tabular}[c]{@{}l@{}}Service Access Control\end{tabular}     & Yes & \cellcolor[HTML]{ffcfcf}No & \cellcolor[HTML]{ffcfcf}$\infty$ & Yes & \cellcolor[HTML]{ffcfcf}No\\ \hline\hline    
 
\parbox[t]{2mm}{\multirow{4}{*}{\rotatebox[origin=c]{90}{Linkerdv2}}} & \begin{tabular}[c]{@{}l@{}}Cluster Message Encryption\end{tabular} & \cellcolor[HTML]{ffcfcf}No & \cellcolor[HTML]{ffcfcf}No  & \cellcolor[HTML]{ffcfcf}N/A & \cellcolor[HTML]{ffcfcf}N/A & \cellcolor[HTML]{ffcfcf}N/A \\ \cline{2-7}
 & \begin{tabular}[c]{@{}l@{}}Service Message Encryption\end{tabular} & Yes & Yes & 24 hours & Yes & Yes \\ \cline{2-7}
 & \begin{tabular}[c]{@{}l@{}}Cluster Access Control\end{tabular} & \cellcolor[HTML]{ffcfcf}No & \cellcolor[HTML]{ffcfcf}No & \cellcolor[HTML]{ffcfcf}N/A  & \cellcolor[HTML]{ffcfcf}N/A & \cellcolor[HTML]{ffcfcf}N/A \\ \cline{2-7}
 & \begin{tabular}[c]{@{}l@{}}Service Access Control\end{tabular} & Yes & \cellcolor[HTML]{ffcfcf}No & \cellcolor[HTML]{ffcfcf}$\infty$** & \cellcolor[HTML]{ffcfcf}No** & \cellcolor[HTML]{ffcfcf}No**\\ \hline\hline   
 
\parbox[t]{2mm}{\multirow{4}{*}{\rotatebox[origin=c]{90}{Istio}}} & \begin{tabular}[c]{@{}l@{}}Cluster Message Encryption\end{tabular} & \cellcolor[HTML]{ffcfcf}No & \cellcolor[HTML]{ffcfcf}No  & \cellcolor[HTML]{ffcfcf}N/A  & \cellcolor[HTML]{ffcfcf}N/A & \cellcolor[HTML]{ffcfcf}N/A \\ \cline{2-7}
 & \begin{tabular}[c]{@{}l@{}}Service Message Encryption\end{tabular} & Yes & \cellcolor[HTML]{ffcfcf}No & Ext Tool~\cite{istio-default-tls20} & Ext Tool~\cite{istio-default-tls20} & Ext Tool~\cite{istio-default-tls20} \\ \cline{2-7}
 & \begin{tabular}[c]{@{}l@{}}Cluster Access Control\end{tabular} & \cellcolor[HTML]{ffcfcf}No  & \cellcolor[HTML]{ffcfcf}No  & \cellcolor[HTML]{ffcfcf}N/A & \cellcolor[HTML]{ffcfcf}N/A & \cellcolor[HTML]{ffcfcf}N/A \\ \cline{2-7}
 & \begin{tabular}[c]{@{}l@{}}Service Access Control\end{tabular} & Yes & \cellcolor[HTML]{ffcfcf}No & \cellcolor[HTML]{ffcfcf}$\infty$** & \cellcolor[HTML]{ffcfcf}No** & \cellcolor[HTML]{ffcfcf}No** \\ \hline\hline
 
\parbox[t]{2mm}{\multirow{4}{*}{\rotatebox[origin=c]{90}{Kubernetes}}} & \begin{tabular}[c]{@{}l@{}}Cluster Message Encryption\end{tabular} & \cellcolor[HTML]{ffcfcf}No* & \cellcolor[HTML]{ffcfcf}No & \cellcolor[HTML]{ffcfcf}N/A & \cellcolor[HTML]{ffcfcf}N/A & \cellcolor[HTML]{ffcfcf}N/A \\ \cline{2-7}
& \begin{tabular}[c]{@{}l@{}}Service Message Encryption\end{tabular} & Yes & \cellcolor[HTML]{ffcfcf}No & 1 year & \cellcolor[HTML]{fbff80}Beta & \cellcolor[HTML]{fbff80}Beta\\ \cline{2-7}
& \begin{tabular}[c]{@{}l@{}}Cluster Access Control\end{tabular}     & Yes & \cellcolor[HTML]{ffcfcf}No & \cellcolor[HTML]{ffcfcf}$\infty$ & \cellcolor[HTML]{ffcfcf}No & \cellcolor[HTML]{ffcfcf}No \\ \cline{2-7}
& \begin{tabular}[c]{@{}l@{}}Service Access Control\end{tabular}     & Yes & \cellcolor[HTML]{ffcfcf}No & \cellcolor[HTML]{ffcfcf}$\infty$ & \cellcolor[HTML]{ffcfcf}No & \cellcolor[HTML]{ffcfcf}No\\ \hline\hline 
 
\end{tabular}
\vspace{.05in}
\caption{Security Mechanisms in Service Mesh Tools -- A summarized view of the security mechanisms 
available in each service mesh tool analyzed, which mechanisms are enabled by default, and additional details about the actual implementations.\\ $^{*}$Pod-to-pod encryption left to third-party implementation~\cite{kube-pod-enc20}. 
\\$^{**}$Inherited from Kubernetes' Role-Based Access Control system~\cite{istio-default-ac20,linkerd-default-ac20}.}
\label{tab:defaults}
\vspace{-0.3in}
\end{table}

Next, we analyze a ``default defender'' in which a service mesh is deployed using the default configuration. 
While service meshes may have security mechanisms available to administrators, we now consider the impact that 
the default configurations have on system security. We frame our experimental evaluation with Consul, 
though we observe similar default behavior among other tools.

\highlight{Consul in Default Configuration}
Consul is unique among the service mesh tools studied in that it provides the functionality of both the service
mesh layer and platform orchestration layer.
Table~\ref{tab:defaults} outlines the available security mechanisms of a
Consul service mesh, along with which of those mechanisms are enabled by
default in the cluster.
While Consul offers all of the necessary security capabilities to
administrators, it {\it fails} to enable any of them by default and lacks rotation support for all mechanisms. With an extensive
list of configurations to create and assign, such as node permissions, key
creation and distribution, and certificate hierarchy, the overhead for system
administrators is significant. 
Within the tutorials Consul provides to their customers, the guidance to
configure the security mechanisms is available, but Consul does not enable
these policies by default and places this onus onto the system
administrators~\cite{consul-acl-tutorial20}.

\highlight{Consul -- Disruption}
Considering a Consul service mesh with default configuration, an adversary is able to take a range of 
actions in order to achieve their goals. Specifically, as previously shown in Figure~\ref{fig:quorum-overwhelm}, 
an adversary is able to flood the Consul quorum with malicious messages disrupting cluster and service availability.
Furthermore, rather than overwhelming the consensus protocol of the cluster, an adversary can simply remove the 
benign members of the cluster through the ``force-leave'' command in the Consul API. 
In order to have access to the Consul API, the adversary must execute a single, unauthenticated command 
once a Consul service mesh cluster is located that allows nodes to join under the default configuration. 
Using the ``force-leave'' command, a malicious user may essentially order cluster members to remove the target node 
from the registry.  Repeating this process for all nodes that the attacker desires to remove, an adversary 
can dismantle the structure of the service mesh and render the availability of the cluster nonexistent.

\highlight{Consul -- Manipulation}
Manipulation of data within a default Consul cluster is trivial once an adversary has joined the cluster.  
The lack of default access control mechanisms within a Consul cluster allows an attacker to modify, create, 
or delete any key/value entries within the cluster.  Also, an adversary may create or read any service 
configurations within the cluster and exfiltrate this information to external nodes for further actions. 
With no default mechanism of limiting reads and writes to the storage, an adversary may observe sensitive data or 
configurations and exfiltrate or modify them as they desire. 
Therefore, under default configurations, sensitive data, or access keys and tokens are entirely vulnerable 
when stored within the key/value storage system of Consul clusters.

\highlight{Consul -- Takeover}
Extending the possibilities of attack under default Consul configuration, is the ability for an adversary to 
assume the leadership position of the cluster.  Due to the election structure of the Consul quorum, combined 
with a lack of access control, new members of the cluster may take any cluster actions, such as issuing the 
``force-leave'' command. Using this command, an adversary may initiate a removal of the current cluster 
leader, introduce new malicious members to the cluster, and attempt to assume the leadership position of the cluster.  
Without any security mechanisms to prevent these actions, an adversary may achieve any goal and any level 
of privilege in a default cluster. In our experimentation a takeover attack was executed by creating a malicious 
node configured to be a Consul server marked with the ``bootstrapper'' flag in order to cause it to request 
a leadership election.  Once this node performed a ``join'' with the cluster, the malicious node and the 
legitimate cluster leader entered a state of conflict where the quorum failed to decide upon the true leader. 
When this state of conflict was reached, the malicious node simply performed the ``force-leave'' 
command on the legitimate node which removed it from the quorum and the benign members adjusted their 
perceived leader to the malicious node.

\highlight{Consul -- Configuration Assessment}
With the default Consul configuration lacking active security mechanisms, the responsibility and burden of 
correctly configuring these mechanisms is placed upon system administrators. The extensive manual tasks required 
to properly secure a cluster create an immense overhead for administrators. Moreover, the likelihood of mistakes 
or errors to occur in large systems due to extensive configurations will be similar to that of large-scale software 
packages~\cite{arcuri08,hangal02}. With adoption of Consul in production environments, supporting and maintaining 
a service mesh infrastructure is a monumental task.  

Aside from the configuration effort of Consul's security mechanisms, the implementation of these mechanisms
demonstrates design inconsistencies with the DevOps ideology.  Consul's base functionality is designed around 
a distributed, decentralized system of connecting microservices.  However, the implementation of its security
mechanisms consolidates control to a singular node and restricts the flexibility of these mechanisms through its lack
of rotation capabilities and lifetime key validity.  While Consul is our primary frame of reference, alternative 
options in Istio and Linkerdv2 present similar difficulties and challenges under their default configurations.

\highlight{Default Configuration Istio and Linkerdv2, with Kubernetes}
As expressed in Section~\ref{sec:back}, Istio and Linkerdv2 both depend upon Kubernetes to 
provide the underlying platform in order to run their service mesh features. Kubernetes is responsible 
for providing the overlaid service mesh with cluster-level message encryption and access control mechanisms. 
However, Kubernetes fails to provide cluster-level message encryption, instead, as shown in Table~\ref{tab:defaults}, 
Kubernetes leaves the implementation of pod-to-pod encryption 
to third-party tools. Also, cluster access control is not implemented by default. 
Due to this, an unknowing system administrator may choose to implement a service mesh such as Istio or Linkerdv2 
with correct configuration of all possible security mechanisms, but be unaware of the underlying vulnerabilities 
of a default Kubernetes. The fragile security structure between these service meshes and Kubernetes
creates a security dependency where tools must implicitly trust one another to provide their claimed functionality. 
With such strong trust required between tools, exploits such as~\cite{seals19,wallen19} show that this 
relationship can be dangerous.

Considering the default configurations of Istio and Linkerdv2, the level of manual configuration
required from system administrators is significant.  As Table~\ref{tab:defaults}
shows, Istio and Linkerdv2 fail to provide means of securing the cluster-level functionality, and service-level 
access control by default.  
However, Linkerdv2 does enable service-to-service message encryption via mutual TLS by default. 
This represents a valuable design decision that adds to the security posture of the overall service mesh. 
In contrast, Istio provides administrators the option of enabling service-to-service message encryption,
but fails to enable this security mechanism by default.  In order to provide the same service-level security,
an administrator would be required to modify configurations of the cluster and provide additional authentication
rules for individual pods and services in order to provide proper, secure functionality.

\section{Related Work}
\label{sec:rel}

To our knowledge, this work is the first systematic study of the security of service mesh tools.
However, many of the classic issues of security, robustness, and functionality are present within service meshes, 
as are several of the attacks that we proposed (albeit many of these attacks are enabled or amplified by novel 
features of service mesh tools). Many of our attacks are inspired by existing work, and many of the implications of our work 
build upon previous studies of microservice security and networked systems. We discuss some of the most closely related work 
to our own below.

\highlight{Microservice Security}
Automation and the decentralized nature of microservice security has been
observed or utilized by a number of previous works. Rastogi,~\etal~\cite{rastogi2017cimplifier} evaluate an automation system 
for dismantling a monolithic software deployment into
a collection of collaborating microservices in order to better adhere to the principle of least privilege~\cite{polp}. Unlike 
our work, Rastogi,~\etal~developed a special-purpose system that uses a static binding to communicate between microservices, rather 
than consider the security of 3rd-party tools that dynamically connect services. Yarygina,~\etal~\cite{yarygina2018overcoming} 
note the comparative lack of security protections for Docker containers, and propose a container security monitor.
In Sun,~\etal~\cite{sun15}, the authors study how the, often inherent, trust relationship between deployed microservices may result in the 
compromise of an entire system.  Further, they propose a system for deploying network security monitors in microservice environments to detect
and block threats to clusters.

A number of previously published works focus on the security of individual Docker containers, which are frequently used as the 
enabling mechanism for individual microservices. 
A representative example is Enck,~\etal~\cite{enckDocker17}, which studies the risk of deploying containers automatically from 
3rd-party container repositories. In Lin,~\etal~\cite{lin18} and Martin,~\etal~\cite{martin18}, the authors examine attacks and countermeasures 
to the security of containers, as well as the ecosystems of repositories and orchestration tools.
Our work assumes that individual containers and repositories are secure, and instead focuses on external threats to the 
mechanisms by which microservices interact. 

\highlight{Analysis of Consensus Protocols}
Some of the attacks that we propose target the RAFT protocol used to form a service mesh.
Some previous work, most notably by Sakic,~\etal~\cite{sakicRAFT2018}, examines the availability and response time of nodes 
participating in RAFT. However, previous work does not consider the influence of an adversary, and is instead concerned 
with the performance of RAFT in a purely-benign setting.

The considerable interest around blockchain technologies has driven the development of security studying microservice clusters 
specifically for running consensus protocols, such as Hyperledger Fabric~\cite{gupta2019peace,sukhwani2017performance}.
These studies observe the threat of sybil attacks on collaborative network services, as does our work. However, blockchain technology 
can defeat traditional sybil attacks via proof-of-work or related protocol-level mechanisms, whereas our attacks require low-latency 
communication and collaboration between microservices.

\highlight{Microservice Attacks} 
The attack vectors that we present are (to our knowledge) unreported. The actual attacks themselves, and the goals of the 
adversaries that we articulate in our threat model are inspired by previous work on attacks against more 
traditional systems. One of the most relevant studies is that of Cherny,~\etal~\cite{chernyBlackhat2017}, which also proposes 
the use of microservice containers as a vector of attacks, thus providing a motivation for services as a target. 
In Csikor,~\etal~\cite{csikor18}, the authors study how specially tailored access control policies crafted 
by an attacker may result in an exhaustion of cloud resources resulting in a denial-of-service to a cluster.


\section{Actionable Advice and Future Work}
\label{sec:future}

Throughout Section~\ref{sec:evaluation}, we note various design flaws and shortcomings
with the current state-of-art in service meshes. We outline these concerns in an effort
to frame future research for service meshes in a direction that incorporates security mechanisms
that more closely align with the goals and benefits that service meshes promise.

\vspace{0.1in}
\highlight{Dynamic and Flexible Security}
Current tools provide various security mechanisms to defend a service mesh, as shown 
in Section~\ref{sec:idealized}. Unfortunately, the design of these mechanism does not align well with 
the desired goals of the service mesh itself. For instance, a TLS-based, centralized security approach 
voids the decentralization benefits provided by the consensus protocols. 
Moreover, some of these security mechanisms are platform-specific and may lock the user into one 
environment e.g., Istio's and Linkeredv2's dependence on Kubernetes. 
In order to synchronize the behaviors of service meshes and the protections afforded through defenses,
the security mechanisms should be part of the core service mesh design. This approach will 
enable greater flexibility and adaptability while aligning the main goals.

\highlight{Synchronized Security Mechanisms}
Illustrated in Table~\ref{tab:attack-table}, the best and most effective security stance for a 
service mesh tool is to have all available security mechanisms enabled and configured in conjunction. 
However, should a security mechanism be forgotten, misconfigured, or exploited over time 
(e.g., obtaining the never-changing UDP key in Consul), areas of exploit emerge immediately 
and the other mechanisms may be rendered futile. To construct better, more secure service 
mesh clusters, it is clear that synchronization and redundancy across security mechanisms is necessary. 
Streamlined security configuration design would provide administrators a simplified experience 
in creating secure clusters.

\highlight{Tailored Security Solutions}
Section~\ref{sec:default} and Table~\ref{tab:defaults} enumerate many shortcomings in the offerings 
of state-of-the-art service meshes.  Many potential attack vectors for service meshes are either
left to third-party tools to implement, or are not enabled by default within the service mesh.  In order
to address the domain-specific security issues for service meshes, security mechanisms and defenses
tailored for this unique area are necessary. 

\section{Conclusions}
\label{sec:conc}
Due to the increase of deployed microservices, service mesh tools appear to be an enticing solution 
to manage and maintain these deployments.  However,
as these tools become more popular and are utilized in production environments, it is necessary 
to assess the available security mechanisms and their strength in deterring adversarial efforts.
As the initial study of service mesh tools used for 
microservice deployments, we examine the three most popular, state-of-art offerings in the service mesh domain
and articulate a threat model tailored to concerns within the service mesh domain.

Through experimentation, we find that under configuration by a skilled administrator, in
10 of the 20 studied scenarios, complete cluster compromise is possible for an attacker.  Further,
in 5 additional scenarios, at least one adversarial goal is achievable.  Under default
configuration, all studied tools, except Linkerdv2, fail to enable 
\underline{any} of their security mechanisms.  These results and our observations in usability
of these mechanisms indicate important design flaws in the security of service 
mesh tools requiring further research and development.

\newpage

\section*{Acknowledgments}
\vspace{-0.05in}
This work is a companion work to that which is set to appear in SecureComm 2020.  The contents of this work are meant to provide extended analyses and additional cases than were provided in the SecureComm article.  The authors would also like to acknowledge Seena Saiedian for their contributions in
proofreading and revising this work.

\bibliographystyle{plain}
\bibliography{100-bibl}

\end{document}